\documentclass[aps,prl,twocolumn,noshowpacs,nofootinbib,superscriptaddress,groupedaddress, floatfix,superscriptaddress, preprintnumbers]{revtex4-1}
 
\usepackage{graphicx} 
\usepackage{dcolumn}   
\usepackage{bm}        

\usepackage[breaklinks,colorlinks=true,linktocpage]{hyperref}

\usepackage{multirow}
\usepackage{feynmf}
\usepackage{amsmath,amssymb}

\usepackage{graphicx,bm}

\usepackage{slashed}

\usepackage{natbib}

\usepackage[usenames]{color}

\usepackage{float}

\newcommand{\Nc}{N_{\rm c}}

\newcommand{\muQ}{\mu_{\rm Q}}
\newcommand{\lqcd}{\Lambda_{\rm QCD}}

\newcommand{\vk}{\boldsymbol{k}}

\newcommand{\vq}{ \boldsymbol{q}}

\newcommand{\calO}{\mathcal{O}}

\newcommand{\rme}{\mathrm{e}}

\newcommand{\up}{\uparrow}
\newcommand{\down}{\downarrow}

\newcommand{\econ}{\tilde{\varepsilon}}

\newcommand{\lamQ}{\lambda_{\rm Q}}
\newcommand{\lamB}{\lambda_{\rm B}}

\newcommand{\fQ}{f_{\rm Q}}
\newcommand{\fB}{f_{\rm B}}
\newcommand{\gQ}{g_{\rm Q}}
\newcommand{\gB}{g_{\rm B}}
\newcommand{\nB}{n_{\rm B}}
\newcommand{\nQ}{n_{\rm Q}}

\newcommand{\chiB}{\chi_{\rm B}}
\newcommand{\muB}{\mu_{\rm B}}
\newcommand{\EB}{E_{\rm B}}
\newcommand{\EQ}{E_{\rm Q}}
\newcommand{\wf}{\varphi}

\newcommand{\MB}{M_{\rm B}}
\newcommand{\MQ}{M_{\rm Q}}
\newcommand{\Lhat}{\hat{L}}
\newcommand{\kF}{k_{\rm F}}
\newcommand{\kFB}{k_{\rm FB}}

\newcommand{\qF}{q_{\rm F}}

\newcommand{\ksat}{k_{\rm sat}}
\newcommand{\kdif}{k_{\rm dif}}

\newcommand{\qsh}{q_{\rm sh}}
\newcommand{\qbu}{q_{\rm bu}}
\newcommand{\ksh}{k_{\rm sh}}
\newcommand{\kbu}{k_{\rm bu}}
\newcommand{\DeltaB}{\Delta_{\rm B}}
\newcommand{\DeltaQ}{\Delta_{\rm Q}}

\newcommand{\Idylliq}{IdylliQ }

\begin{document}
\begin{flushright}
\end{flushright}
\preprint{INT-PUB-23-018}

\title{Momentum Shell in Quarkyonic Matter from Explicit Duality:
\\
 A Dual Model for Cold, Dense QCD}

\author{Yuki~Fujimoto}
\email{yfuji@uw.edu}
\affiliation{Institute for Nuclear Theory, University of Washington, Box 351550, Seattle, WA, 98195, USA}

\author{Toru~Kojo}
\email{torujj@nucl.phys.tohoku.ac.jp}
\affiliation{Department of Physics, Tohoku University, Sendai 980-8578, Japan}

\author{Larry~D.~McLerran}
\email{mclerran@me.com}
\affiliation{Institute for Nuclear Theory, University of Washington, Box 351550, Seattle, WA, 98195, USA}

\date{\today}

\begin{abstract}
We present a model of cold QCD matter that bridges nuclear and quark matter through the duality relation between quarks and baryons.
The baryon number and energy densities are expressed as functionals of either the baryon momentum distribution, $\fB$, or the quark distribution, $\fQ$,
which are subject to the constraints on fermions, $0 \le f_{\rm B,Q} \le 1$.
The theory is ideal in the sense that the confinement of quarks into baryons is reflected in the duality relation between $\fQ$ and $\fB$,
while other possible interactions among quarks and baryons are all neglected.
The variational problem with the duality constraints is formulated and we explicitly construct analytic solutions,
finding two distinct regimes: 
A nuclear matter regime at low density and a Quarkyonic regime at high density. 
In the Quarkyonic regime, baryons underoccupy states at low momenta but form a momentum shell with $\fB=1$ on top of a quark Fermi sea. 
Such a theory describes a rapid transition from a soft nuclear equation of state to a stiff Quarkyonic equation of state. 
At this transition, there is a rapid increase in the pressure.
\end{abstract}


\maketitle

\textit{Introduction.}---Understanding cold, dense QCD matter is a difficult problem.
Slightly above nuclear saturation density, the importance of many-body forces complicates the physical picture.
The distinction between baryon and quark degrees of freedom is not clear-cut and in fact a proper description should allow a \emph{dual} simultaneous description of both quarks and baryons.
This paper attempts such a dual description.

We construct such a dual model on the basis of very simple principles.
This model is analytically solvable.
It has its consequences that the high-density phase is \emph{Quarkyonic}.
The notion of Quarkyonic matter has emerged from studies of cold, dense QCD in the limit of a large number of colors, $\Nc \rightarrow \infty$,
where the quark screening effects to color confinement are suppressed by a factor $1 / \Nc$.
At a finite quark chemical potential $\muQ$,
the confinement persists to $\muQ \sim \sqrt{\Nc} \lqcd$ ($\lqcd \simeq 300$ MeV: dynamical scale in QCD) until quarks Debye screen gluons,
while quarks establish the Fermi sea at much lower density, with $\muQ \sim \lqcd$.
At $\lqcd \ll \muQ \ll \sqrt{\Nc}\lqcd$, there must be  quark matter with the confinement.
This paradoxical feature was resolved by assuming baryons as effective degrees of freedom on top of a quark Fermi sea~\cite{McLerran:2007qj} (see, however, Ref.~\cite{Koch:2022act}).
This dual feature, with quarks and baryons in different domains of momenta for a single phase of matter,
is suitable to describe continuous evolution from nuclear matter to weakly-coupled quark matter~\cite{Freedman:1976xs,*Freedman:1976ub, Kurkela:2009gj, Gorda:2018gpy, Gorda:2021znl,*Gorda:2021kme}.

Models of Quarkyonic matter are shown to reveal an equation of state (EOS) that rapidly stiffens from low to high density~\cite{McLerran:2018hbz, Fukushima:2015bda, Jeong:2019lhv, Duarte:2020xsp,*Duarte:2020kvi,  Zhao:2020dvu, Margueron:2021dtx} (see also Refs.~\cite{Philipsen:2019qqm, Sen:2020peq, Cao:2020byn, Kovensky:2020xif, Duarte:2021tsx,*Duarte:2023cki})
and therefore has features consistent with the observed properties of neutron stars~\cite{Demorest:2010bx, Antoniadis:2013pzd, Fonseca:2016tux, NANOGrav:2017wvv, NANOGrav:2019jur, Fonseca:2021wxt, LIGOScientific:2017vwq,*LIGOScientific:2018hze,*LIGOScientific:2018cki, Riley:2019yda, Miller:2019cac, Miller:2021qha, Riley:2021pdl,Raaijmakers:2021uju}.
Notably, these features encompass a rapid rise in the sound speed~\cite{Masuda:2012kf,*Masuda:2012ed, Kojo:2014rca, Bedaque:2014sqa, Tews:2018kmu,Fujimoto:2017cdo,*Fujimoto:2019hxv,*Fujimoto:2021zas, Drischler:2020fvz, Kojo:2020krb, Drischler:2021bup, Kojo:2021wax, Altiparmak:2022bke, Brandes:2022nxa} and the vanishing of the trace anomaly signifying the conformal nature of dense matter~\cite{Fujimoto:2022ohj, Marczenko:2022jhl, Ma:2018xjw,Lee:2021hrw, Ecker:2022dlg, Annala:2023cwx} (see also Ref.~\cite{Chiba:2023ftg}).

In this work, we present dynamical descriptions for Quarkyonic matter by making full use of the duality between baryons and quarks~\cite{Kojo:2021ugu, Kojo:2021hqh} (see also Refs.~\cite{Ma:2019xtx, Ma:2021zev}).
The baryon and energy densities are expressed by either baryonic or quark degrees of freedom,
characterized by the occupation probability of momentum states, $\fB(k)$ and $\fQ(q)$, for baryons and quarks, respectively
(the letters $k$ and $q$ are used exclusively for baryon and quark momentum, respectively).
Since baryons are composed of quarks, there is a relationship between $\fB$ and $\fQ$,
allowing a dual description without double counting of quark contributions.

\begin{figure}
  \centering
  \vspace{-0.5cm}
  \includegraphics[width=0.95\columnwidth]{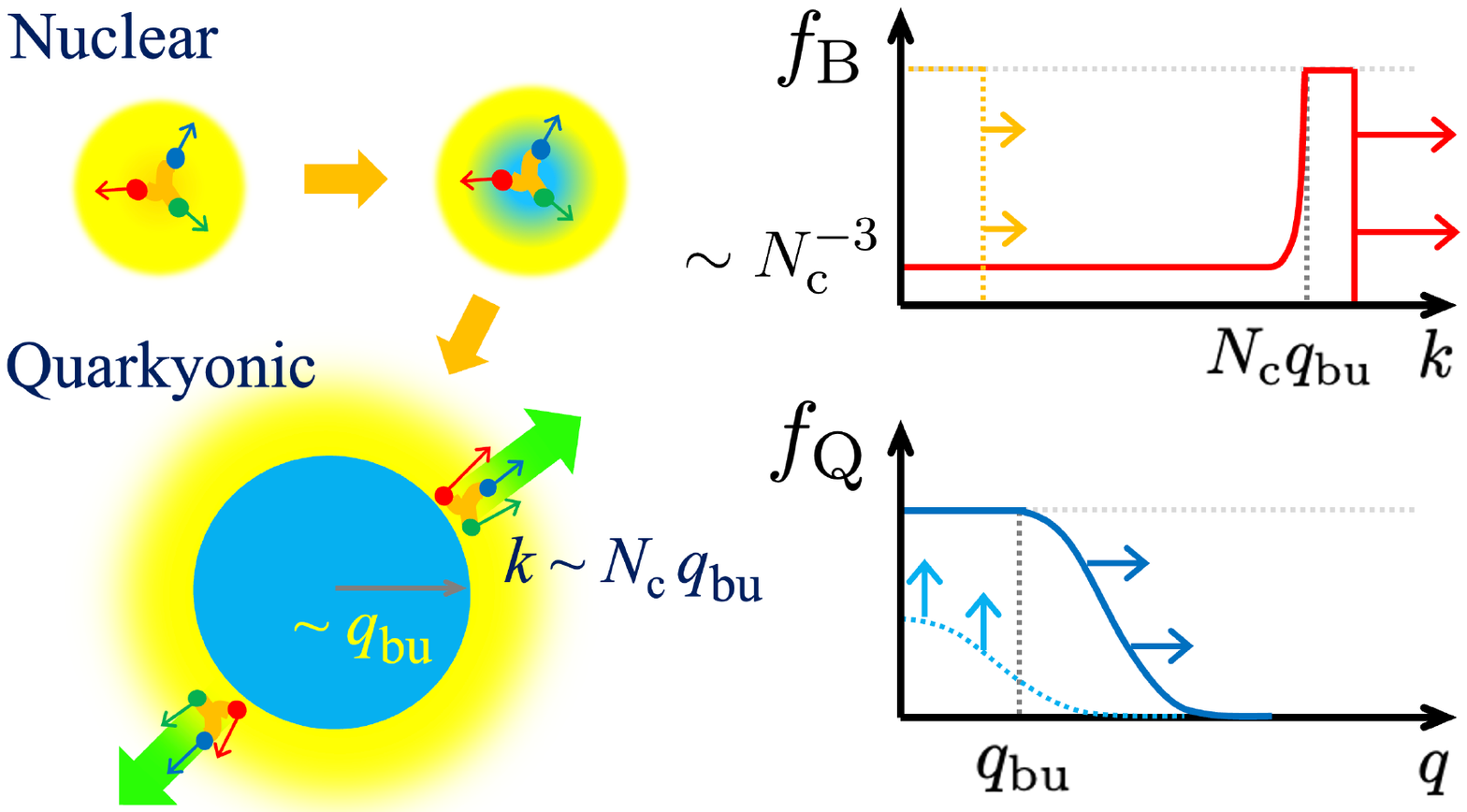}
    \vspace{-0.5cm}
  \caption{Evolution of $\fB$ and $\fQ$ from the nuclear (dotted) to Quarkyonic regime (solid).
    The saturation of quark states drives baryons into the relativistic regime.
    Arrows in the rightmost panels indicate increasing $\muB$.
  }
    \vspace{-0.5cm}
  \label{fig:shell}
\end{figure}

The simplest description one can imagine of such matter is that except for the confinement of quarks into baryons, interactions of quarks and baryons can be ignored.
Such an idealized matter we will call \emph{\Idylliq} (Ideal Dual Quarkyonic) matter.
We formulate a variational problem to minimize the energy density functional $\varepsilon_{\rm B} [\fB] = \varepsilon_{\rm Q} [\fQ]$ for a given baryon density $\nB$,
and determine the distribution $\fB$ and $\fQ$.
Even such an idealized model is nontrivial due to the duality relation and quantum mechanical constraints $0 \le f_{\rm Q, B}  \le 1$.
The quark substructure constraint is crucial for rapid evolution of stiffness from nuclear to quark matter~\cite{Kojo:2021ugu, Kojo:2021hqh}.
For this \Idylliq model, we establish the momentum shell of baryons on top of the quark Fermi sea as schematically shown in Fig.~\ref{fig:shell}.
\Idylliq theory give a novel alternative explanation for the previously proposed picture of Quarkyonic matter~\cite{McLerran:2007qj,McLerran:2018hbz}.

\vspace{0.5em}

\textit{Duality.}---We posit the duality relation between $\fQ$ for
quarks with {\it a given color} and $\fB$ for baryons as (notation: $\int_{k} \equiv \int \frac{d^d \vk}{(2\pi)^d} $)~\cite{Kojo:2021ugu, Kojo:2021hqh}.
\begin{equation}
  \label{eq:duality}
  \hspace{-0.3cm}
  \big[ \fQ (q) \big]_{f\sigma} 
  =\!\!\!\! \sum_{ i=n,p,\cdots} \sum_{\sigma'=\up,\down}  \int_k \bigg[ \wf\big(\vq - \tfrac{\vk}{\Nc}\big) \bigg]_{f\sigma}^{i \sigma'} \!\!\!\! \big[\fB(k)\big]_{i \sigma'}\,,
\end{equation}
where $\wf$ is a single quark momentum distribution with the flavor $f$ and spin $\sigma$ in a single baryon state of a species $i$ and spin $\sigma'$.
Collecting quark contributions from each baryon leads to quark distributions in dense matter.
In this work, we limit ourselves to  symmetric nuclear matter and include a spin-isospin degeneracy factor 4 in the expressions of thermodynamic quantities,
but elsewhere we drop  the spin-flavor indices $f,\sigma$. 
The extension for multi-flavors and multi-baryon species will be discussed in the forthcoming papers.

The normalization is $\int_q \wf(q) = 1$.
The dual expression of the baryon number readily follows from Eq.~\eqref{eq:duality} as
\begin{equation}
  \label{eq:duality_n_e}
   \nB = 4\int_k \fB(k) = 4 \int_q \fQ(q) \,.
\end{equation}
The energy densities in terms of baryons and quarks are
\begin{equation}
  \begin{split}
    \varepsilon_{\rm B} [\fB] &= 4 \int_k \EB(k) \fB(k)  \,,\\
    \varepsilon_{\rm Q} [\fQ] &= 4 \int_q \EQ(q) [\Nc \fQ(q)] \,.
\end{split}
\end{equation}
Remember $\fQ$ is defined for a fixed color, $\fQ \equiv \fQ^{R} = \fQ^{G} = \fQ^{B} $ with which $\nB=\nQ^R=\nQ^G=\nQ^B$.
A single baryon is assumed to have the energy contributions summed from $\Nc$-confined quarks, 
$\EB (k) = \Nc \int_q \EQ (q) \wf \big( \vq - \vk/\Nc \big) $.
Then a duality relation follows, $\varepsilon = \varepsilon_{\rm B} [\fB] = \varepsilon_{\rm Q} [\fQ]$.
As quarks are confined in a spatial domain of the baryon size $\sim \lqcd^{-1}$,
quarks can be energetic and $\wf (q) $ is spread to momenta of $\sim \lqcd$.
The \emph{mechanical} pressure inside of a baryon is large.

In this work, going from low to high densities we keep using the same $\wf$ determined in vacuum.
Our main target here is the transient regime from baryonic to quark matter,
where using $\wf$ for localized quarks may not be so bad approximation.
The structural changes in baryons, such as swelling, would possibly increase the low momentum components of $\wf$,
but such modifications merely shift the onset of quark matter formation to lower density.

\vspace{0.5em}

\textit{Minimization of energy functional.}---With duality~\eqref{eq:duality} as a constraint, 
we calculate the energy density $\varepsilon$ for a given $\nB$.
We consider energy functionals
\begin{equation}
\varepsilon = \varepsilon_{\rm B} [\fB] \big|_{\nB } = \varepsilon_{\rm Q} [\fQ] \big|_{\nB } \,,
\end{equation}
and minimize them by optimizing $\fB$ or $\fQ$ while holding $\nB$ fixed.
A novelty in our optimization program is that
the solutions are determined not only by the stationary condition $\delta \varepsilon/\delta f =0$
but also by the boundary conditions $f_{\rm B, Q}=0$ or 1.
The thermodynamic energy density is obtained by substituting the optimized distributions,
$\varepsilon_{\rm EOS} (\nB) =  \varepsilon_{\rm B} [\fB^*]  \big|_{\nB } = \varepsilon_{\rm Q} [\fQ^*]  \big|_{\nB } $.

In practice, one can find the $\fB^*$ and $\fQ^*$ by minimizing
\begin{equation}
  \econ = \varepsilon_{\rm B} [\fB] - \lamB \nB =  \varepsilon_{\rm Q} [\fQ] - \lamQ \nQ \,,
\end{equation}
where $\lamB = \Nc \lamQ$.
It is tempting to identify the $\lambda$'s as chemical potentials and $\econ$ as the thermodynamic functional.
Unfortunately they do not satisfy the thermodynamic relations 
if solutions are partly determined by the boundary conditions.
Hence we use $\econ$ only to find $\fB^*$ and $\fQ^*$, and use them in computations of $\varepsilon_{\rm EOS} (\nB)$.

\vspace{0.5em}

\textit{Global constraints.}---The constraints in our theory appear {\it global},
as $\fQ$ at a given momentum depends on $\fB$ for the entire momentum range.
The variation leads to
\begin{equation}
\frac{\delta \econ}{\delta \fB(k)} = \EB(k) - \lamB\,,\quad
\frac{\delta \econ}{\delta \fQ(q)} = \EQ(q) - \lamQ\,.
\end{equation}
At momenta with $\delta \econ/\delta f_{\rm B,Q} < 0 $, greater $f_{\rm B,Q}$ reduces $\econ$ 
and grows toward the boundary $f_{\rm B,Q} =1$,
while $\delta \econ/\delta f_{\rm B,Q} > 0 $ drives $f_{\rm B,Q}$ to the other boundary, $f_{\rm B,Q} =0$.
We would get the optimized distributions
\begin{equation}
\fB^{\rm var} (k) = \Theta (\kF-k)\,,\quad \fQ^{\rm var}  (q) = \Theta (\qF-q)\,,
\label{eq:step}
\end{equation}
where $\kF$ and $\qF$ are determined through $\lamB = \EB(\kF)$ and $\lamQ = \EQ(\qF)$.

The above solutions are not usable everywhere.
For instance, $\fQ^{\rm var}$ at large momenta is incompatible with the sum rule~\eqref{eq:duality};
at large momenta ($q\gg k/\Nc$), the scaling should be $\fQ (q) \sim \nB \wf(q)$.
Another problem is that, if we keep using $\fB^{\rm var}$ in the regime $ \lqcd \ll \kF \ll \Nc \lqcd$,
then $\fQ (0) \sim \nB \wf(0) \sim \kF^3/\lqcd^3$, violating $\fQ \le 1$ at $q=0$.
Our problem is to patch the candidates of solutions,
found from variational calculations and the boundary conditions, 
into the form consistent with the duality constraints,
and then to minimize the energy.

\vspace{0.5em}

\textit{Solvable model.}---What makes the dual theory nontrivial is its global nature.
The difficulty lies in the reconstruction of $\fB$ from a given $\fQ$.
At high density we have good reasoning to choose $\fQ = 1$ for some interval of $q$.
But it is difficult to tell which $\fB$ gives $\fQ=1$ while not violating $\fQ \le 1$ anywhere.
To uncover the general features of the dual theory,
we choose a specific $\wf$ which reduces the global problem 
to the one determining a couple of global constants.
We choose 
\begin{equation}
  \label{eq:wf3d}
  \wf_{\rm 3d}(\vq) = \frac{2\pi^2}{\Lambda^3} \frac{e^{-q/\Lambda}}{q/\Lambda}\,,
\end{equation}
which is the inverse of a linear differential operator
$ \Lhat = - \nabla_q^2 + \frac1{\Lambda^2}$.
Applying this operator to the sum rule~\eqref{eq:duality}, 
we find the local relation between $\fB$ and $\fQ$ ($d=3)$,
\begin{equation}
  \label{eq:derivrel}
  \fB(\Nc q) =  \frac{\Lambda^2}{\Nc^d} \, \Lhat \big[ \fQ(q) \big] \,.
\end{equation}
Here we assume $\EB(k) = \sqrt{k^2 + \MB^2}$ with $\MB$ being a constant as we focus on the deconfining aspect.
The single quark energy can also be determined as
\begin{equation}
  \label{eq:EQ}
  \hspace{-0.0cm}
    \EQ = \sqrt{q^2 + \MQ^2 \,}\, \bigg(1- \frac{ (d-1)\Lambda^2}{q^2 + \MQ^2} - \frac{\MQ^2 \Lambda^2}{(q^2 + \MQ^2)^{2}} \bigg)\,,
\end{equation}
where $\MQ \equiv \MB / \Nc$.
We note $\EQ' (q) > 0 $ everywhere.

We must examine which $\fQ$ satisfies $\fB=0$ or $\fB=1$.
For this purpose we introduce $y_\pm (q) = \rme^{\pm q/\Lambda}/q$
which satisfies $\Lhat [y_\pm] = 0$.
The boundary $\fB(\Nc q) = 0$ can be obtained as $\fQ^{\fB=0} (q) = c_+ y_+ (q) + c_- y_- (q) $.
Meanwhile $\fB (\Nc q) = 1$ can be obtained as $\fQ^{\fB=1} (q) = \Nc^d + d_+ y_+ (q) + d_- y_- (q) $.
The constants $c_\pm$ and $d_\pm$ must be chosen to keep $0 \le \fQ(q) \le 1$.

Now we have exhausted candidates of local solutions for $\fQ$; the boundary values $\fQ=0,1$ and those dual to $\fB=0,1$.
The question is how to patch them.
At momenta where two different solutions meet,
the second derivative in $\Lhat$ and the condition $0\le \fB\le 1$ demands $\fQ$ to be continuous up to the first derivative.
For example, acting $\Lhat$ on a function $\fQ^{\rm bu} = \eta(q) \Theta (\qbu-q)$ 
generates the terms $\eta' (\qbu) \delta (q - \qbu)$ and $\eta ( \qbu ) \delta'(q - \qbu)$. 
To cancel such delta's violating the condition $\fB \le 1$, we have to add a function $\fQ^{\rm joint} = \xi(q) \Theta (q-\qbu)$
with $\xi(\qbu) = \eta(\qbu)$ and $\xi' (\qbu) = \eta' (\qbu)$.
We construct such a $\fQ^{\rm joint}$ using solutions $\fQ^{\fB=1} $ and $\fQ^{\fB=0}$.

\vspace{0.5em}

\begin{figure*}
  \centering
    \vspace{-0.3cm}
  \includegraphics[width=0.99\textwidth]{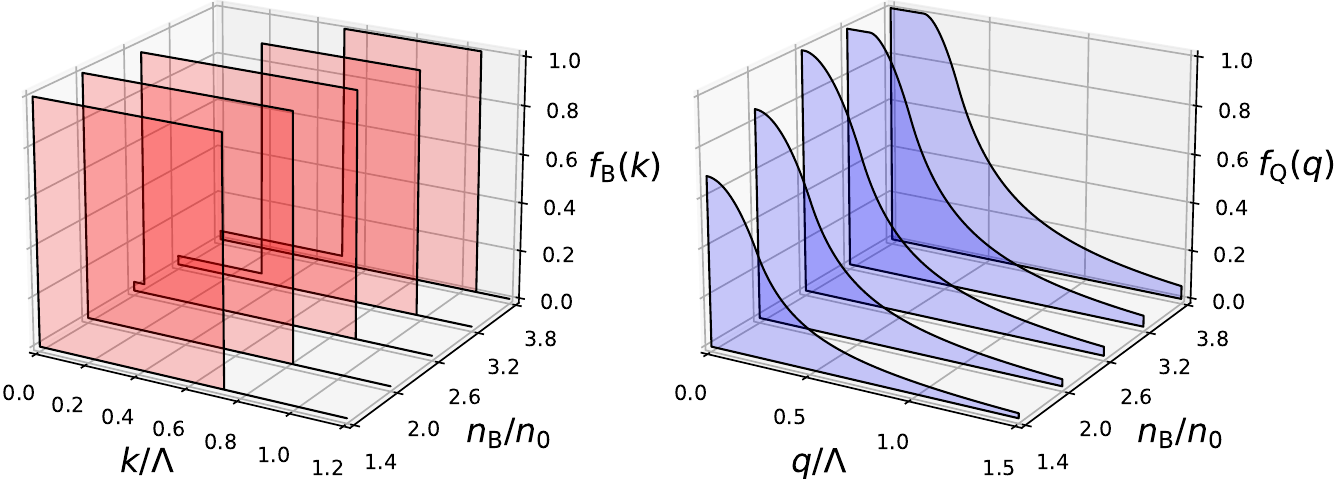}
  \caption{Evolution of $\fB(k)$ (left) and $\fQ(q)$ (right) with increasing $\nB$ for $\Lambda=0.4\ \text{GeV}$.
  }
    \vspace{-0.2cm}
  \label{fig:evol}
\end{figure*}

\textit{Transitions from baryonic to quark matter.}---We go over from a dilute baryonic matter to a dense quark matter by patching the candidates of local solutions.

In dilute matter one can simply use the ideal baryon gas $\fB^{\rm idB} (k) = \Theta (\kF - k)$ and its dual expression $\fQ^{\rm idB} $.
This regime continues until $\fQ^{\rm idB}$ reaches the upperbound.
It occurs first at $q=0$ when $ \kF \simeq \sqrt{2/\Nc} \Lambda$
or $ \nB /n_0 = 2.58 \times \big( \Lambda /0.4\,\text{GeV} \big)^3$ for $\Nc =3$ ($n_0 \simeq 0.16~\text{fm}^{-3}$:  normal nuclear density).
Note that this implies the saturation density is parametrically small compared to the QCD scale $\Lambda^3$.

In the post-saturation regime, we can no longer use the ideal baryon gas picture.
For the low momentum part of $\fQ$, the only candidate is $\fQ=1$.
We found that 
the solution must involve three segments (two segment models cannot satisfy the continuity at the first derivative),
\begin{align}
    \fQ(q) = \Theta (\qbu - q )
    &+ \fQ^{\fB=1} (q) \Theta (\qsh - q )\Theta (q - \qbu ) \notag \\
    &+ \fQ^{\fB=0} (q) \Theta (q - \qsh ) \,,
  \label{eq:fQpostsat}
\end{align}
where $d_+$ in $\fQ^{\fB=0}$ must be zero.
Its dual baryon distribution is ($k=\Nc q, \kbu =\Nc \qbu, \ksh=\Nc \qsh)$
\begin{equation}
\fB(k) 
  = \frac{1}{\Nc^{d} } \, \Theta (\kbu - k ) +\Theta (\ksh - k )\Theta (k - \kbu ) \,,
  \label{eq:fBpostsat}
\end{equation}
which is small in the \emph{bulk} Fermi sea at $k \leq \kbu$ but forms the baryon \emph{shell} with the maximum height at $\kbu < k \leq \ksh$,
reproducing the form conjectured by McLerran and Reddy \cite{McLerran:2018hbz}.
This shape is energetically favored as $\fB$ and $\fQ$ are kept as compact as possible.
Figure~\ref{fig:evol} shows the forms of $\fB^{\rm idB}$ and $\fQ^{\rm idB}$ in the dilute regime at $\nB / n_0 \lesssim 2.6$, and the forms of $\fB$~\eqref{eq:fBpostsat} and $\fQ$~\eqref{eq:fQpostsat} in the post-saturation regime at $\nB / n_0 \gtrsim 2.6$.

With four conditions from two junction points, one can express $c_\pm$, $d_-$,
and $\Delta_{\rm Q} = \qsh - \qbu$ as functions of $\qsh$.
The parameter $\qsh$ is determined by observing that
the $\delta \econ/\delta \fB > 0 $ for $\EB > \lamQ$ 
introduces the energy cost unless
$\fB$ drops from 1 to 0 at $\lamB = \EB(\Nc \qsh)$.
Here we display only the equation to determine $\Delta_{\rm Q}$
as it is needed for computations of EOS.
The equation to be solved is
\begin{equation}
  \label{eq:hkFB}
   \frac{\Lambda + \qbu}{\Lambda + \qbu - (\Lambda + \qsh) e^{-\DeltaQ / \Lambda}} = \Nc^3 \,.
\end{equation}
Below, we discuss the thickness of the baryon momentum shell, $\DeltaB = \Nc \DeltaQ$, which is obtained as a solution of this transcendental equation.
Close to the momentum $\ksh = \ksat$ at which $\fQ$ saturates, we expand the equation with respect to $\delta k \equiv \ksh - \ksat $, then the solution is approximately
\begin{equation}
  \DeltaB \simeq \ksh - \sqrt{\frac{2\ksat}{1 + \ksat / (\Nc \Lambda)} \delta k}\,,
  \label{eq:deltasingular}
\end{equation}
where we assumed $\delta k \ll \Lambda$ and $\kbu \ll \Lambda$ then expanded them up to $\calO(\delta k)$ and $\calO(\kbu^2)$, respectively.
The $\ksh$-derivative of $\DeltaB \propto \sqrt{\delta k}$ diverges as 
$\delta k^{-1/2} $ for $\delta k\rightarrow 0$.

At large momentum, $\ksh > \Lambda$, the shell thickness is
\begin{equation}
\DeltaB \simeq \frac{\Lambda}{\Nc^2} + \frac{\Lambda^2}{ \Nc \ksh} \,.
\end{equation}
It approaches constant for a large $\ksh$.

\begin{figure}
  \centering
  \includegraphics[width=0.95\columnwidth]{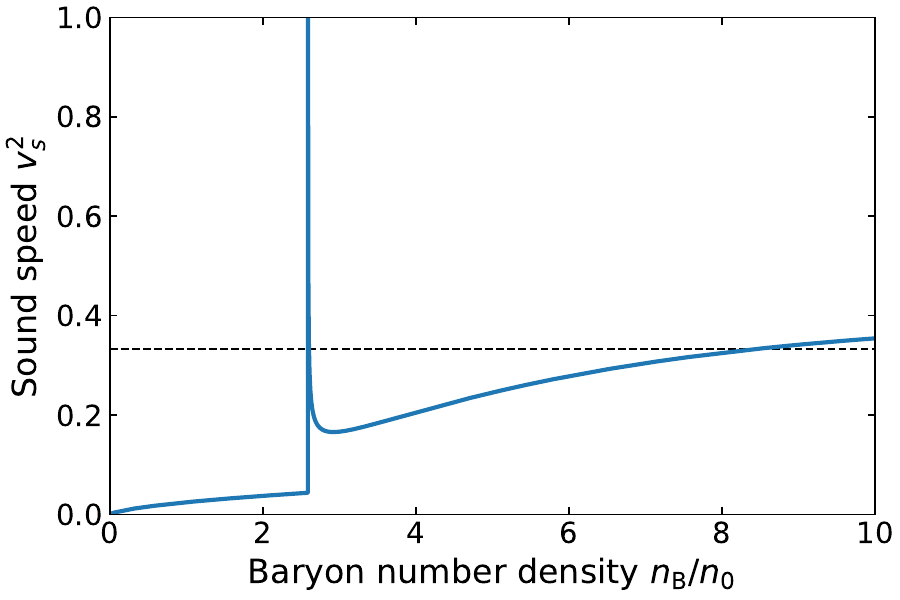}
  \caption{Sound speed $v_s^2$ as a function of $\nB/n_0$ for $\Lambda = 0.4\ \text{GeV}$.}
  \vspace{-0.2cm}
  \label{fig:vs2}
\end{figure}

\vspace{0.5em}

\textit{Equations of state.}---We examine the unified EOSs.
Below  quark saturation, the EOSs are simply those of the ideal baryon gas,
$\nB^{\rm below} = 2\kF^3/3\pi^2$ and
$\varepsilon^{\rm below} = 4 \int_k \EB (k) \Theta(k - \kF)$.
Above the saturation, the bulk part of $\fB$ is depleted,
\begin{equation}
  \begin{split}
\nB^{\rm above} 
&= 4 \! \int_{\kbu}^{\ksh }\!\!\!\! \frac{d^3 \vk}{(2\pi)^3} 
  + \frac{4}{\Nc^3} \!\int_0^{\kbu} \!\!\frac{d^3 \vk}{(2\pi)^3} \,,\\
\varepsilon^{\rm above}
&= 4 \! \int_{\kbu}^{\ksh } \!\!\!\! \frac{d^3 \vk}{(2\pi)^3} \EB(k)
  + \frac{4}{\Nc^3} \! \int_0^{\kbu} \!\!\!\! \frac{d^3 \vk}{(2\pi)^3} \EB(k)\,.
\end{split}
  \label{eq:above_sat}
\end{equation}
Due to the depletion, the growth of $\ksh$ increase $\nB$ more slowly than  in the pre-saturation regime,
but the energy per particle $\varepsilon/ \nB$ grows much faster in the post-saturation regime.    
Accordingly the pressure $P = \nB^2 \partial (\varepsilon/\nB)/\partial \nB$ is large;
the EOS is stiff.

It is important to examine whether thermodynamic quantities are continuous at quark saturation.
At saturation $\kbu=0$ in Eq.~\eqref{eq:above_sat} so that $\nB$ and $\varepsilon$ are continuous.
Next we check whether derivatives of $\varepsilon$ with respect to $\nB$ are continuous. 
We first compute
\begin{equation}
2\pi^2 \frac{\partial \nB^{\rm above} }{\partial \ksh}
= \ksh^2 - \bigg(1 - \frac{1}{\Nc^3} \bigg) \kbu^2 \frac{\partial \kbu }{\partial \ksh} \,.
\end{equation}
Because of the phase space factor $\kbu^2$, at  saturation the second term specific to the post-saturation regime vanishes,
leaving continuous $\partial \nB/\partial \ksh$.
Similarly $\partial \varepsilon /\partial \ksh$ is continuous and so are $\muB = \partial \varepsilon/\partial \nB$ and the pressure $P=\muB \nB-\varepsilon$.
Note that this continuity, relying on the vanishing phase space for $\kbu\rightarrow 0$, does not hold in 1+1 dimensions; 
indeed a 1+1 dimensional \Idylliq model yields discontinuous $\muB$ which is not permitted in the thermodynamics.

In 3+1 dimensional \Idylliq theory, unfortunately the continuity holds only up to the first derivative.
The baryon susceptibility $\chiB$ has a discontinuity at  saturation and so does the sound speed $v_s^2$.
The susceptibility $\chiB$ drops discontinuously;  this dropping should not be confused with that in a second order phase transition where $\chiB$ jumps up.
Figure~\ref{fig:vs2} shows the behavior of $v_s^2$ as a function of $\nB/n_0$ for $\Lambda = 0.4 \ \text{GeV}$.
Note that in our model, $v_s^2$ may exceed the conformal value $v_s^2 = 1/3$ even at high densities where we expect it to be subconformal~\cite{Freedman:1976xs,*Freedman:1976ub, Kurkela:2009gj, Gorda:2018gpy, Gorda:2021znl,*Gorda:2021kme}, depending on the value of $\Lambda$.
Also, as mentioned earlier, $v_s^2$ is singular at saturation and the parametric dependence of the singular part, $\hat{v}_s^2$, on $\delta k$ is
\begin{equation}
  \label{eq:vs2sing}
   \hat{v}_s^2
  \sim - \frac{\ksh \ksat}{\MB^2} \frac{d\DeltaB}{d\ksh} \sim \frac1{\Nc^{3}} \sqrt{\frac{\ksat}{\delta k}}\,,
\end{equation}
given that $\ksh \sim \ksat \sim \Nc^{-1/2} \Lambda$ and $\MB \sim \Nc \Lambda$.

\vspace{0.5em}

\textit{Minimal corrections to \Idylliq model.}---We outline how the singular behavior of $\DeltaB$ and $v_s^2$ is remedied by smoothing out the sharp edge of the baryon distribution.
As the divergent part of $v_s^2$~\eqref{eq:vs2sing} is proportional to $d\DeltaB / d \ksh$, we focus on the singular behavior of $\DeltaB$.
We define a function $\gQ(k)$, which is the quark occupation at the origin corresponding to the baryon Fermi sea filled up to momentum $k$:
\begin{equation}
  \gQ(k) \equiv
  \int_{k'} \wf\left(\frac{k'}{\Nc} \right) \Theta(k - k')\,.
  \label{eq:gQ}
\end{equation}
Below the saturation density, $\kFB = \ksat - 0^+$, 
the condition $\fQ(q=0) \rightarrow 1$ is equivalent to $\gQ(\kFB) \rightarrow 1$.

Above the saturation density, the condition $\fQ(q=0)=1$ becomes
\begin{equation}
  \gQ(\ksh) - \bigg(1 - \frac1{\Nc^3} \bigg) \gQ(\kbu) = 1\,,
  \label{eq:gksat}
\end{equation}
At saturation, this equation is equivalent to Eq.~\eqref{eq:hkFB} that sets the relation between $\ksh$ and $\DeltaB$.

By taking the $\ksh$-derivative on both sides of Eq.~\eqref{eq:gksat}, we obtain
\begin{equation}
  \label{eq:dDdksh}
  \frac{d \DeltaB(\ksh)}{d\ksh} = - \frac{\Nc^3}{\Nc^3 - 1} \frac{\gQ'(\ksh)}{\gQ'(\kbu)} + 1\,,
\end{equation}
where $\gQ'(k) = e^{-k /(\Nc\Lambda)} \times \Nc k/\Lambda^2$. 
Near the saturation, $\gQ'(\kbu) \sim \kbu \propto \delta k^{1/2} \rightarrow 0$ yields the singularity in $d \DeltaB(\ksh) / d\ksh$ and $\hat{v}_s^2$.
The problematic $\gQ'(k) \propto k$ scaling comes from the $\partial \Theta(k-k')/\partial k = \delta(k-k')$ term
which picks out the integrand $\sim k'^2 \varphi(k'/\Nc) \sim k'$ {\it exactly} at $k'=k$.
A little smearing of the baryon Fermi surface cures this problem:
replacing $\Theta(k - k')$ with a smooth function $\gB(k - k')$ whose damping scale is $\sim \kdif$, 
one can make $\gQ' (k) \sim \kdif$ finite for $k\rightarrow 0$.
In turn, at saturation we have $d \DeltaB(\ksh) / d\ksh \sim - \ksh/\kdif$ and $ \hat{v}_s^2 \sim \MB^{-2} \ksat^3/\kdif \sim \Nc^{-3} \ksat/\kdif$.
For the causal sound speed, the required width is $\kdif \gtrsim \Nc^{-3} \ksat$, much smaller than the Fermi momentum at saturation.
We note that too large smearing washes out the peak structure in $v_s^2$ by reducing the disparity between the nuclear and quark pressure. 
Viewing nuclear forces as quark exchanges would explain the precursor behavior toward the quark regime.
Leaving aside the details of such smoothing,
our theory firmly establishes an inevitable stiffening caused by the quark substructure.

\vspace{0.5em}

\textit{Summary and discussions.}---The description \Idylliq matter we present is ultimately very simple.
At low densities, there is a filled Fermi distribution of nucleons, and quarks may be thought of as degrees of freedom inside the nucleons with momenta out to $\sim \lqcd$.
At some density there is a transition characterized by the saturation of quark states.
At higher density, quarks form a filled Fermi sea with an exponentially falling tail above some momentum.
In the dual description, baryons under-occupy a bulk Fermi sea but form a fully filled shell at a Fermi surface.  
These distributions are shown in Fig.~\ref{fig:evol}.

We chose the specific model~\eqref{eq:wf3d} for $\wf$ to solve the \Idylliq theory exactly.
The following findings should be universal for other choices of $\varphi$:
(a) The saturation of the quark distribution $\fQ$, as demonstrated for different $\wf$~\cite{Kojo:2021ugu, Kojo:2021hqh},
and the underoccupation of $\fB$ at lower momenta coming from the quark substructure constraint;
(b) The asymptotic behavior of $\fQ(q) \sim \nB \wf(q)$ at $q\rightarrow \infty$; 
(c) Existence of the shell structure in $\fB$, energetically favored to make $\fQ$ a more compact distribution.
Actually these properties largely survive for the \Idylliq model perturbed by wide class of interaction functionals of $f_{\rm B,Q}$.
Hence the \Idylliq model offers a good baseline for theories of Quarkyonic matter.

\vspace{0.5em}

\begin{acknowledgments}
  We thank
  Gordon Baym,
  Jean-Paul Blaizot,
  Kenji Fukushima,
  Tetsuo Hatsuda,
  Masakiyo Kitazawa,
  Rob Pisarski,
  Sanjay Reddy,
  and
  Volodya Vovchenko
  for useful conversations.
  Y.F.\ is supported by the Japan Society for the Promotion of Science (JSPS) through the Overseas Research Fellowship.  
  The work of Y.F.\ and L.M.\ was supported by the U.S. DOE under Grant No. DE-FG02-00ER41132.
  The work of T.K.\ was supported by JSPS KAKENHI Grant No. 23K03377 and by the Graduate Program on Physics for the Universe (GPPU) at Tohoku university.
\end{acknowledgments}

\bibliographystyle{apsrev4-1}
\bibliography{bib_shell}

\end{document}